\documentclass[sigplan,nonacm]{acmart}
\usepackage{pgfplots}
\usepackage{xcolor}
\usepackage{siunitx}
\usepackage{booktabs}
\usepackage{tabularx}
\pgfplotsset{compat=1.18}
\usetikzlibrary{shapes.geometric, arrows.meta, positioning, external}

\usepackage{algorithm}
\usepackage{algpseudocode}
\usepackage{amsmath}

\begin{document}

\title{Affinity Tailor: Dynamic Locality-Aware Scheduling at Scale}

\author{Jin Xin Ng}
\email{njx@google.com}
\affiliation[obeypunctuation=true]{%
  \institution{Google},
  \country{USA}
}

\author{Ori Livneh}
\email{atdt@google.com}
\affiliation[obeypunctuation=true]{%
  \institution{Google},
  \country{USA}
}

\author{Richard O'Grady}
\email{rjogrady@google.com}
\affiliation[obeypunctuation=true]{%
  \institution{Google},
  \country{USA}
}

\author{Josh Don}
\email{joshdon@google.com}
\affiliation[obeypunctuation=true]{%
  \institution{Google},
  \country{USA}
}

\author{Peng Ding}
\email{dingp@google.com}
\affiliation[obeypunctuation=true]{%
  \institution{Google},
  \country{USA}
}

\author{Samuel Grossman}
\email{samuelgr@google.com}
\affiliation[obeypunctuation=true]{%
  \institution{Google},
  \country{USA}
}

\author{Luis Otero}
\email{lotero@google.com}
\affiliation[obeypunctuation=true]{%
  \institution{Google},
  \country{USA}
}

\author{Chris Kennelly}
\email{ckennelly@google.com}
\affiliation[obeypunctuation=true]{%
  \institution{Google},
  \country{USA}
}

\author{David Lo}
\email{davidlo@google.com}
\affiliation[obeypunctuation=true]{%
  \institution{Google},
  \country{USA}
}

\author{Carlos Villavieja}
\email{villavieja@google.com}
\affiliation[obeypunctuation=true]{%
  \institution{Google},
  \country{USA}
}

\renewcommand{\shortauthors}{Affinity Tailor: Dynamic Locality-Aware Scheduling at Scale}

\begin{abstract}
Modern large multicore systems often run multiple workloads that share CPUs under schedulers such as Linux CFS. To keep CPUs busy, these schedulers load-balance runnable work, causing each workload to execute on many cores. This weakens locality at the microarchitectural level: workloads lose reuse in caches, branch predictors, and prefetchers, and interfere more with one another—especially on chiplet-based systems, where spreading execution across cores also spreads it across LLC boundaries. A natural alternative is strict CPU partitioning, but hard partitions leave capacity idle when workloads do not fully use their reserved CPUs. 

We present \textit{Affinity Tailor}, a userspace-guided kernel scheduling system built on a key insight: the kernel can preserve locality for workloads that share CPUs by treating demand-sized, topologically compact CPU sets as affinity hints rather than hard partitions. A userspace controller estimates each workload’s CPU demand online and assigns a preferred CPU set sized to that demand, chosen to be as disjoint as possible from other workloads while spanning as few LLC domains as possible. The kernel then uses this set as an affinity hint, steering threads toward those CPUs while still allowing execution elsewhere when needed to preserve utilization. Deployed at Google, Affinity Tailor delivers geometric-mean per-CPU throughput gains of 12\% on chiplet-based systems and 3\% on non-chiplet systems over Linux CFS. Furthermore, faster execution reduces memory residency, yielding per-GB throughput gains of 3-7\%. Our findings suggest that future schedulers should treat spatial locality as a first-class objective, even at the expense of work-conservation.
\end{abstract}

\maketitle

\section{Introduction}
As modern datacenter processors aggressively scale core counts, individual applications increasingly struggle to saturate massive hardware topologies. To maximize hardware utilization, hyperscalers aggressively co-locate hundreds of workloads per machine \cite{bashir2021take,baset2012towards}, resulting in operating system schedulers continuously interleaving threads from distinct applications on the same physical cores. This interleaving is inefficient due to the continuous loss of microarchitectural state. Every time the kernel schedules a new thread onto a core, it effectively begins diluting the previous thread's instruction execution and data access history, thus degrading the hit rates across the memory hierarchy (L1, L2, and last-level caches) and reducing the accuracy of core-level predictive hardware structures, such as the branch predictor and hardware prefetchers. Concurrently, memory bandwidth per core has stagnated in recent server architectures, while workloads have become memory-intensive \cite{jain2024limoncello}. This disparity leaves shared memory interconnects highly vulnerable to saturation caused by traffic from antagonistic workloads.

Thread scheduling is a primary determinant of application performance \cite{lozi2016linux}. Modern OS schedulers—such as Linux's \textit{Completely Fair Scheduler} (CFS) \cite{molnar2007cfs} and \textit{Earliest Eligible Virtual Deadline First} (EEVDF) scheduler \cite{stoica1996eevdf, zijlstra2023eevdf}—prioritize \textit{work-conservation}. A strictly work-conserving scheduler operates on the axiom that no CPU core should sit idle if a runnable thread exists. Consequently, the OS constantly multiplexes threads across cores to minimize immediate queueing. While recent advances have made scheduling policies more extensible \cite{humphries2021ghost, sched_ext2024}, work-conservation remains a dominant paradigm.

Work-conserving approaches maximize theoretical CPU cycle utilization but are increasingly at odds with modern hardware. Over the last several generations of server CPUs, the cadence of Moore's law has driven core counts exponentially higher while memory bandwidth per core has stagnated. On modern processor architectures composed of numerous \textit{chiplet-based} core complexes, non-uniform cache hierarchies pose significant inefficiencies for application performance as memory traffic must traverse shared memory interconnects \cite{zhou2024characterizing}, often with fixed bandwidth ceilings. Even when the bandwidth ceilings are not reached, elevated memory bandwidth results in higher memory latency \cite{jain2024limoncello}. Simultaneously, software applications spend a significant portion of cycles \textit{backend-bound} \cite{accelerometer,kanev2015profiling}, i.e., waiting for data to load. When a work-conserving scheduler migrates a thread between processor cores, it forces cache lines to move across the processor, further congesting shared interconnects and degrading the performance of all co-located workloads.

Operating systems provide mechanisms like \texttt{cpusets} to strictly limit thread execution to specific cores. However, static partitioning is ill-suited for modern datacenters for two reasons. First, workloads are notoriously bursty; strict limits cause severe latency penalties during spikes in parallelism. Second, datacenter operators frequently \textit{overcommit} \cite{bashir2021take,baset2012towards} resources---the sum of CPU resources sold on a machine exceeds the physical machine capacity---making it impossible to assign each application a disjoint set of CPUs. Thus, while \texttt{cpusets} can preserve microarchitectural state, they are fundamentally incompatible with the bursty, overcommitted environment of modern datacenters.

In the literature, hardware-assisted resource partitioning mechanisms \cite{lo2015heracles,chen2019parties,zhang2013cpi2,patel2020clite,chen2023olpart} have been discussed as a means of managing the use of shared system resources between co-located applications, but these systems do not govern thread-to-core placement. Separately, the Nest scheduler \cite{lawall2022scheduling} concentrates threads onto a dense set of warm cores to exploit higher turbo frequencies, but lacks the notion of per-application isolation. Cache Aware Scheduling patches \cite{chen2024cas} in the Linux kernel propose LLC-affinity heuristics in the load balancer, but operate with minimal insight into applications' execution histories. None of these systems provide the dynamic, locality-aware, application-isolating scheduling needed in modern warehouse-scale datacenters.

We introduce Affinity Tailor, an OS scheduling architecture that opportunistically maximizes spatial locality without strictly sacrificing work-conservation. Affinity Tailor introduces a novel Linux kernel mechanism, \textit{Preferred Cores}, to provide \textit{soft affinity}. Threads are drawn to dynamically sized, "hot" execution domains during nominal load, thus improving spatial locality and better retaining microarchitectural state. Specifically, Affinity Tailor promotes thread execution on "hot" preferred cores—featuring primed caches, prefetchers, and an accurate branch predictor. The Preferred Cores mechanism acts as a permeable boundary, permitting threads to burst onto external cores during momentary parallelism spikes to prevent excessive thread queuing. 

Affinity Tailor utilizes two distinct core allocation strategies. We initially developed a chiplet-granularity algorithm for split-LLC architectures. By applying soft affinity to pack applications into the absolute minimum number of required chiplets, this algorithm sought to reduce cross-boundary migrations to alleviate shared interconnect saturation. Our initial fleet deployments revealed unexpected efficiency gains in per-core predictive structures—specifically L1/L2 caches and branch predictors—independent of the LLC. Driven by these insights, we developed a secondary, fine-grained core allocation algorithm tailored specifically for monolithic-LLC architectures. 

Unlike previously proposed microsecond-scale scheduling systems \cite{kaffes2019shinjuku, ousterhout2019shenango, fried2020caladan, iyer2023concord, jia2024skyloft, lin2024fast, li2024libpreemptible} that replace the OS scheduler with custom scheduling stacks, Affinity Tailor functions transparently within the kernel to maximize spatial locality, and is compatible with arbitrary server hardware and software applications.

We deployed and evaluated both algorithms of Affinity Tailor across thousands of machines in Google's global fleet over one week across a highly diverse mix of latency-sensitive user-facing services and throughput-oriented batch workloads. Our evaluation demonstrates that Affinity Tailor improves aggregate application throughput by up to 12\% per-CPU and up to 7\% per-GB memory,  enabling systems to harvest most of the performance benefits of isolated scheduling domains without sacrificing work-conservation.

In summary, this paper makes the following key contributions:
\begin{itemize}
    \item We present Preferred Cores, a novel Linux kernel mechanism providing cgroup-based soft affinity that preserves work-conservation.
    \item We introduce a userspace system that dynamically sizes soft affinity regions using short-horizon demand predictions.
    \item We deployed Affinity Tailor in Google's production fleet, evaluating it across four distinct server platforms and observing aggregate application throughput improvements of up to 12\% per-CPU and up to 7\% per-GB memory on highly utilized machines.
    \item We empirically demonstrate that aggressively load-balancing threads to prevent immediate queueing—the foundational principle of modern OS schedulers—is actively detrimental on modern hardware and software architectures.
\end{itemize}

\section{Background and Motivation}\label{sec:motivation}
Affinity Tailor is motivated by three converging hardware and operational trends: the economic realities of warehouse-scale computing, the constraints of modern multi-core topologies, and the limitations of existing operating system isolation mechanisms.

\subsection{Economics of Overcommitment}\label{sec:motivation:overcommit}
Hyperscalers operate under tight economic constraints: energy, power, and physical hardware are scarce resources \cite{barroso2018datacenter, lo2015heracles, csis2024bottleneck}. Furthermore, organic growth in computational demand routinely outpaces the physical supply of newly procured hardware.

\begin{figure}[htbp] 
    \centering
    \includegraphics{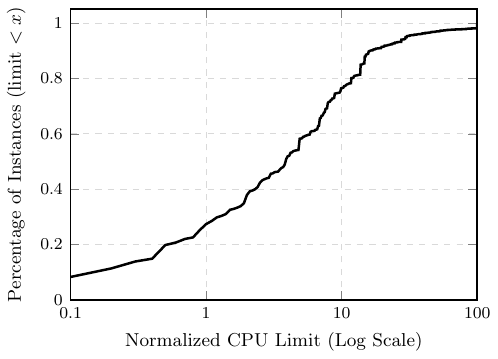}
    \caption{The proportion of application instances executing at or below a given normalized CPU limit in Google's fleet. CPU limits are normalized to represent equivalent computational power across server generations.}
    \label{fig:limit_cdf_log}
\end{figure}

While modern datacenter processors now feature massive topologies with 256 logical cores per socket, individual application sizes have largely not followed this trend. Some workloads simply cannot scale proportionally due to Amdahl's Law, as vertical scaling is ultimately bottlenecked by serial computation. However, many others are restricted by operational constraints, such as strict reliability and failover requirements that favor distributed instances, or by the necessity of provisioning fragmented capacity to handle diurnal traffic patterns \cite{verma2015borg, tirmazi2020borg}. Figure \ref{fig:limit_cdf_log} shows that the vast majority of applications in Google's fleet request fewer than 10 CPUs.

\begin{figure}[htbp]
\centering
\begin{tikzpicture}
\begin{axis}[
    xlabel={Year (Server Generations)},
    ylabel={Logical Cores},
    xmin=2016.5, xmax=2025.5,
    ymin=0, ymax=300,
    x tick label style={/pgf/number format/1000 sep={}},
    xtick={2017, 2018, 2019, 2020, 2021, 2022, 2023, 2024, 2025},
    ymajorgrids=true,
    grid style=dashed,
    nodes near coords={\pgfplotspointmeta},
    point meta=explicit symbolic, 
    every node near coord/.append style={anchor=west, font=\scriptsize},
    width=\columnwidth,
    height=0.65\columnwidth,
    only marks %
]

\addplot[color=black!70, mark=*]
    coordinates {
        (2017.54,56)  [56]
        (2019.29,56)  [56]
        (2021.29,72)  [72]
        (2023.04,112) [112]
        (2023.96,128) [128]
        (2024.71,256) [] %
    };

\addplot[color=black!70, mark=*]
    coordinates {
        (2019.63,128) [128]
        (2021.21,128) [128]
        (2022.88,192) [192]
        (2024.79,256) [256]    %
    };

\end{axis}
\end{tikzpicture}
\caption{Logical CPU cores per socket across recent server generations, showing a 4.6x increase from 56 to 256 CPUs over the last eight years.}
\end{figure}
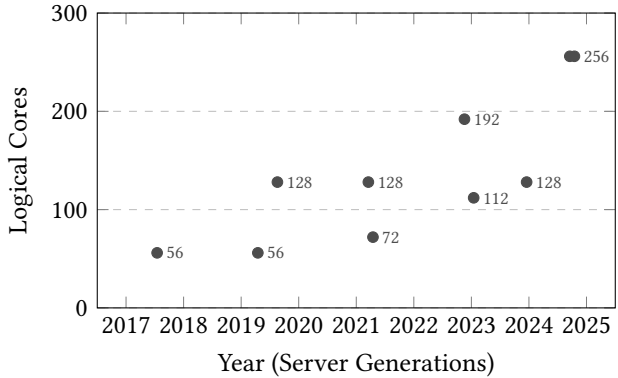

To bridge this massive core-count vs. application size gap and minimize Total Cost of Ownership (TCO), cluster managers like Google's Borg \cite{verma2015borg} must rely on aggressive multi-tenancy. Dedicated servers are only economically viable in limited circumstances such as truly latency-critical workloads, as they come with a price premium. For general services, hundreds of disparate workloads are co-located onto the same physical server to maximize hardware utilization. Cluster managers employ a technique known as \textit{overcommitment} \cite{bashir2021take, baset2012towards}  to ensure even higher load factors---allowing the sum of \textit{requested} CPUs to frequently exceed the sum of \textit{physical} CPUs on the system. Statistical models are used to ensure that individual applications are likely to have access to their requested CPUs. Crucially, high machine utilization exacerbates underlying physical bottlenecks, such as in memory bandwidth saturation and overheads through the loss of microarchitectural state.

\subsection{Microarchitectural Interference}
The aggressive multi-tenancy required by modern economics is fundamentally at odds with the trajectory of modern hardware architectures. The most critical structural bottleneck exacerbated by overcommitment is memory bandwidth. Over the last several generations of server CPUs, the ratio of memory bandwidth available per CPU core has stagnated.

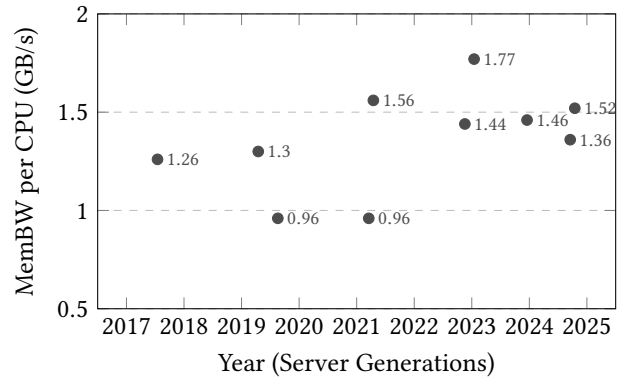
\begin{figure}[htbp]
\centering
\begin{tikzpicture}
\begin{axis}[
    xlabel={Year (Server Generations)},
    ylabel={MemBW per CPU (GB/s)},
    xmin=2016.5, xmax=2025.5,
    ymin=0.5, ymax=2, %
    x tick label style={/pgf/number format/1000 sep={}},
    xtick={2017, 2018, 2019, 2020, 2021, 2022, 2023, 2024, 2025},
    ymajorgrids=true,
    grid style=dashed,
    nodes near coords,
    nodes near coords style={/pgf/number format/fixed, /pgf/number format/precision=2},
    every node near coord/.append style={anchor=west, font=\scriptsize},
    width=\columnwidth,
    height=0.65\columnwidth,
    only marks
]

\addplot[color=black!70, mark=*]
    coordinates {
        (2017.54,1.26) %
        (2019.29,1.30) %
        (2021.29,1.56) %
        (2023.04,1.77) %
        (2023.96,1.46) %
        (2024.71,1.36) %
    };

\addplot[color=black!70, mark=*]
    coordinates {
        (2019.63,0.96) %
        (2021.21,0.96) %
        (2022.88,1.44) %
        (2024.79,1.52) %
    };

\end{axis}
\end{tikzpicture}
\caption{Memory bandwidth per logical core across recent server generations. Relative to the rise in core counts, memory bandwidth per core has remained comparatively flat while per-core compute efficiency has generally improved.}
\end{figure}

This degradation is compounded by operating systems' ineffective preservation of spatial locality. When an OS scheduler such as CFS strictly adheres to work-conservation, it aggressively load-balances threads across the entire processor socket to prevent immediate thread queueing. While theoretically optimal for cycle utilization, this constant thread migration induces severe microarchitectural interference:

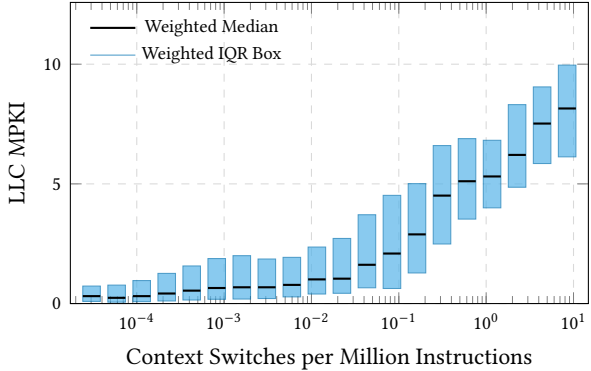
\begin{figure}[t]
\centering
\begin{tikzpicture}
\definecolor{boxcolor}{HTML}{56B4E9}    %
\definecolor{edgecolor}{HTML}{0072B2}   %
\definecolor{mediancolor}{HTML}{000000} %
\begin{axis}[
    width=\linewidth,
    height=2.2in,
    xmode=log,
    log basis x={10},
    xlabel={Context Switches per Million Instructions},
    ylabel={LLC MPKI},
    xmin=1.736701e-05, xmax=1.480045e+01,
    ymin=0, ymax=12.5800,
    grid=major,
    grid style={dashed, gray!30},
    legend style={at={(0.02,0.98)}, anchor=north west, font=\scriptsize, draw=none, fill=none},
    tick label style={font=\scriptsize},
    label style={font=\small},
]

    \draw [fill=boxcolor, draw=edgecolor, opacity=0.7] (2.41115e-05, 0.0900) rectangle (3.82626e-05, 0.7300);
    \draw [mediancolor, thick] (2.41115e-05, 0.3100) -- (3.82626e-05, 0.3100);

    \draw [fill=boxcolor, draw=edgecolor, opacity=0.7] (4.66901e-05, 0.0600) rectangle (7.40926e-05, 0.7700);
    \draw [mediancolor, thick] (4.66901e-05, 0.2400) -- (7.40926e-05, 0.2400);

    \draw [fill=boxcolor, draw=edgecolor, opacity=0.7] (9.04118e-05, 0.0800) rectangle (0.000143475, 0.9600);
    \draw [mediancolor, thick] (9.04118e-05, 0.3100) -- (0.000143475, 0.3100);

    \draw [fill=boxcolor, draw=edgecolor, opacity=0.7] (0.000174874, 0.1100) rectangle (0.000277508, 1.2600);
    \draw [mediancolor, thick] (0.000174874, 0.4200) -- (0.000277508, 0.4200);

    \draw [fill=boxcolor, draw=edgecolor, opacity=0.7] (0.000337851, 0.1500) rectangle (0.000536136, 1.5700);
    \draw [mediancolor, thick] (0.000337851, 0.5400) -- (0.000536136, 0.5400);

    \draw [fill=boxcolor, draw=edgecolor, opacity=0.7] (0.000653469, 0.1800) rectangle (0.00103699, 1.8800);
    \draw [mediancolor, thick] (0.000653469, 0.6500) -- (0.00103699, 0.6500);

    \draw [fill=boxcolor, draw=edgecolor, opacity=0.7] (0.00126539, 0.1900) rectangle (0.00200805, 2.0000);
    \draw [mediancolor, thick] (0.00126539, 0.6800) -- (0.00200805, 0.6800);

    \draw [fill=boxcolor, draw=edgecolor, opacity=0.7] (0.00244751, 0.2100) rectangle (0.00388396, 1.8600);
    \draw [mediancolor, thick] (0.00244751, 0.6800) -- (0.00388396, 0.6800);

    \draw [fill=boxcolor, draw=edgecolor, opacity=0.7] (0.00473396, 0.2800) rectangle (0.00751234, 1.9300);
    \draw [mediancolor, thick] (0.00473396, 0.7800) -- (0.00751234, 0.7800);

    \draw [fill=boxcolor, draw=edgecolor, opacity=0.7] (0.00915641, 0.4000) rectangle (0.0145303, 2.3600);
    \draw [mediancolor, thick] (0.00915641, 1.0100) -- (0.0145303, 1.0100);

    \draw [fill=boxcolor, draw=edgecolor, opacity=0.7] (0.0177103, 0.4300) rectangle (0.0281044, 2.7200);
    \draw [mediancolor, thick] (0.0177103, 1.0400) -- (0.0281044, 1.0400);

    \draw [fill=boxcolor, draw=edgecolor, opacity=0.7] (0.0342551, 0.6600) rectangle (0.0543594, 3.7100);
    \draw [mediancolor, thick] (0.0342551, 1.6200) -- (0.0543594, 1.6200);

    \draw [fill=boxcolor, draw=edgecolor, opacity=0.7] (0.066256, 0.6300) rectangle (0.105142, 4.5200);
    \draw [mediancolor, thick] (0.066256, 2.0900) -- (0.105142, 2.0900);

    \draw [fill=boxcolor, draw=edgecolor, opacity=0.7] (0.128152, 1.2800) rectangle (0.203364, 5.0100);
    \draw [mediancolor, thick] (0.128152, 2.8900) -- (0.203364, 2.8900);

    \draw [fill=boxcolor, draw=edgecolor, opacity=0.7] (0.247871, 2.4900) rectangle (0.393346, 6.6000);
    \draw [mediancolor, thick] (0.247871, 4.5100) -- (0.393346, 4.5100);

    \draw [fill=boxcolor, draw=edgecolor, opacity=0.7] (0.47943, 3.5300) rectangle (0.760808, 6.8900);
    \draw [mediancolor, thick] (0.47943, 5.1100) -- (0.760808, 5.1100);

    \draw [fill=boxcolor, draw=edgecolor, opacity=0.7] (0.92731, 4.0000) rectangle (1.47155, 6.8200);
    \draw [mediancolor, thick] (0.92731, 5.3100) -- (1.47155, 5.3100);

    \draw [fill=boxcolor, draw=edgecolor, opacity=0.7] (1.7936, 4.8600) rectangle (2.84626, 8.3100);
    \draw [mediancolor, thick] (1.7936, 6.2100) -- (2.84626, 6.2100);

    \draw [fill=boxcolor, draw=edgecolor, opacity=0.7] (3.46917, 5.8500) rectangle (5.50522, 9.0500);
    \draw [mediancolor, thick] (3.46917, 7.5200) -- (5.50522, 7.5200);

    \draw [fill=boxcolor, draw=edgecolor, opacity=0.7] (6.71004, 6.1300) rectangle (10.6482, 9.9600);
    \draw [mediancolor, thick] (6.71004, 8.1500) -- (10.6482, 8.1500);

\addlegendimage{color=mediancolor, thick}\addlegendentry{Weighted Median}
\addlegendimage{fill=boxcolor, draw=edgecolor, opacity=0.7}\addlegendentry{Weighted IQR Box}
\end{axis}
\end{tikzpicture}
\caption{Impact of context switching rates on last-level cache misses per-kilo-instruction (LLC MPKI) in Google's fleet. The trend demonstrates that as context switching rates increase, LLC MPKI increases, highlighting the penalties of frequent thread re-scheduling.}
\label{fig:trendmpki}
\end{figure}

\textbf{Core-Level Predictive Structures}
The most immediate consequence of aggressive thread migration is the continuous pollution of core-local predictive state. Every time a thread is migrated, it leaves behind its "warm" execution history. As the thread begins executing on a new, "cold" remote core, it suffers a steep increase in branch mispredictions and L1/L2 cache misses while simultaneously evicting the state of the previous tenant.

Furthermore, modern datacenter processors depend on highly tuned cache replacement and prefetching algorithms to mask latency \cite{jain2024limoncello}. Aggressive thread migration effectively blinds these predictive mechanisms; without a stable execution history, their accuracy drops, leading to degradations in overall processor efficiency. Additionally, individual cores are forced to handle diverse working sets from distinct workloads, further degrading the efficacy of these predictive structures. 

\textbf{Last-Level Cache (LLC)}
Beyond private caches, thread migration degrades the shared LLC. As demonstrated in Figure \ref{fig:trendmpki}, we observe a clear correlation between increased context switching rates and higher LLC misses per-kilo-instruction. This is particularly severe in chiplet-based designs where the LLC is split into discrete slices. When a thread is load-balanced across a chiplet boundary, its previously established LLC cache lines are effectively orphaned, forcing the hardware to fetch data from remote cache slices.

\textbf{Memory Bandwidth} As migrating threads suffer local-LLC misses due to degraded LLC performance, they trigger a flood of cross-chiplet, and main-memory reads. In modern architectures, constant data movement saturates the shared memory interconnects, starving all co-located tenants of the already scarce per-core memory bandwidth. 

Memory bandwidth congestion is severely compounded by hardware prefetchers, which aggressively issue memory fetches in an attempt to hide data access latency. As recent work such as Limoncello \cite{jain2024limoncello} has demonstrated, in bandwidth-constrained environments, aggressive hardware prefetching becomes actively detrimental; modern systems must dynamically disable these prefetchers under load to improve overall throughput. Conversely, minimizing cross-chiplet data movement inherently reduces memory interconnect saturation, thus allowing hardware prefetchers to remain active and effective for longer durations.

\subsection{Hard Affinity \& Bursting}
To combat microarchitectural interference, one seemingly obvious solution is to enforce strict execution isolation via the Linux kernel's native \texttt{cpuset} subsystem. This imposes \textit{hard affinity}, strictly binding threads to rigid CPU boundaries. However, these static solutions categorically fail at the scale of modern datacenters for two primary reasons.

First, hard affinity is mathematically incompatible with aggressive resource overcommitment. In overcommitted environments where the aggregate requested CPU limits exceed physical machine capacity, it is impossible to assign disjoint CPU sets to all tenants without violating the pigeonhole principle. Attempting to overlap hard affinity boundaries forces arbitrary contention on shared CPU. Because workloads peak at unpredictable intervals, co-locating tightly-bound workloads risks causing \textit{throughput violations} due to severe thread queueing. Workloads could be starved of requested CPU resources while cores outside the overlapping boundary remain idle. Furthermore, relying on userspace cluster agents to resolve these violations by altering the assigned CPU sets is ineffective; their second-scale reaction times are incapable of mitigating microsecond-scale workload bursts.

\begin{figure}
  \centering
  \includegraphics[width=\columnwidth]{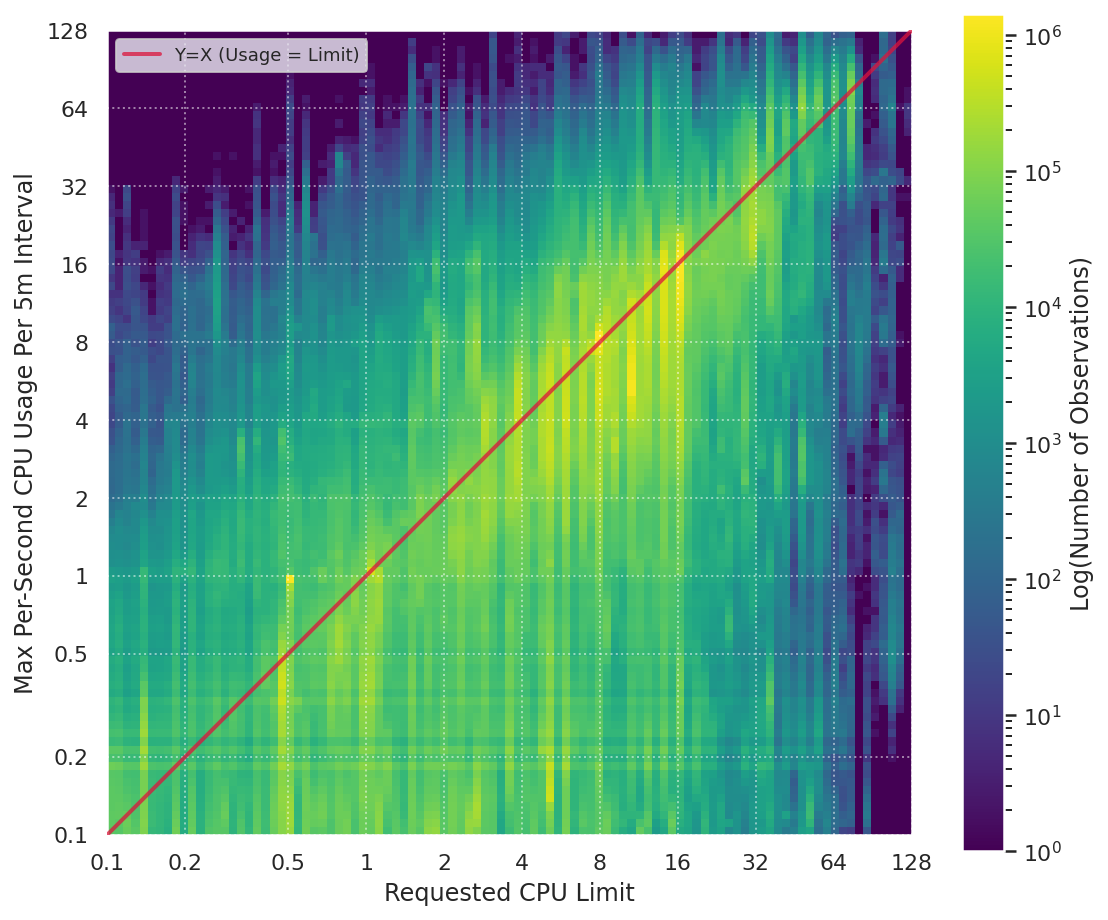}
  \caption{Density map of maximum per-second CPU usage within each 5-minute interval versus requested CPU limit, showing that workloads frequently burst well beyond their nominal CPU allocations.}
  \label{fig:parallelism_heatmap}
\end{figure}

Second, even in the absence of overcommitment, strict CPU limits are hostile to microsecond-scale traffic bursts. Fleet telemetry from Google's production servers in Figure \ref{fig:parallelism_heatmap} indicates that individual applications routinely rely on bursting well beyond their requested CPU limits. Unpredictable spikes in parallelism quickly saturate strictly bounded CPU sets. Under hard affinity, this saturation results in immediate, localized thread queueing, generating unacceptable tail-latency spikes for user-facing services. To accommodate these bursty workloads without stranding capacity, operating systems seeking to improve spatial locality require dynamic, permeable boundaries rather than rigid jails.

\section{System Architecture}\label{sec:system_architecture}

Affinity Tailor comprises three components: a kernel mechanism that implements preferential thread steering, a demand predictor that estimates short-term CPU requirements, and a topologically-aware core allocation algorithm that assigns disjoint affinity regions to the containers on a given machine.

\subsection{The Preferred Cores Kernel Mechanism}
\textit{Preferred Cores} is a novel Linux kernel mechanism which operates on a per-cgroup basis, providing a cgroupfs interface to specify a set of preferred CPUs. Unlike \texttt{cpusets},  which impose \textit{hard affinity} by strictly restricting thread placement, Preferred Cores introduces a \textit{soft affinity} heuristic. The scheduler favors these cores to maximize locality but retains the flexibility to use non-Preferred Cores.

The custom scheduling mechanism is built into a component that we named Core-Aware Scheduling (CAS). CAS was inserted into the thread wakeup and load-balancing paths in the Linux scheduler. When a thread becomes runnable, CAS evaluates placement using a tiered approach:
\begin{itemize}
    \item \textbf{Fast path (preferred cores scan):} CAS intersects the container's runnable \texttt{cpuset} with its dynamically assigned Preferred Cores mask. It scans this subset for an idle\footnote{We additionally use a set of heuristics to identify a core which is likely to become idle soon, in similar fashion to CFS.} core. If found, the thread is enqueued there immediately.
    \item \textbf{Slow path (runnable cores scan):} If an idle core is not found in the fast path, i.e., if capacity within the Preferred Cores mask is fully saturated, CAS falls back to scanning the broader runnable \texttt{cpuset} for an idle core.
\end{itemize}

During load-balancing, CAS prevents threads queued within Preferred Cores from being migrated to non-Preferred Cores, and aggressively migrates both running and runnable threads towards idle Preferred Cores at regular intervals.

The design of CAS strictly preserves work-conservation, as runnable threads are still permitted to use any available CPU, irrespective of the configured Preferred Cores. Notably, if all containers can be given disjoint sets of Preferred Cores, threads of distinct containers would be strongly "attracted" to those disjoint regions, thereby minimizing spillover into the shared CPU regions and maximizing spatial locality. 

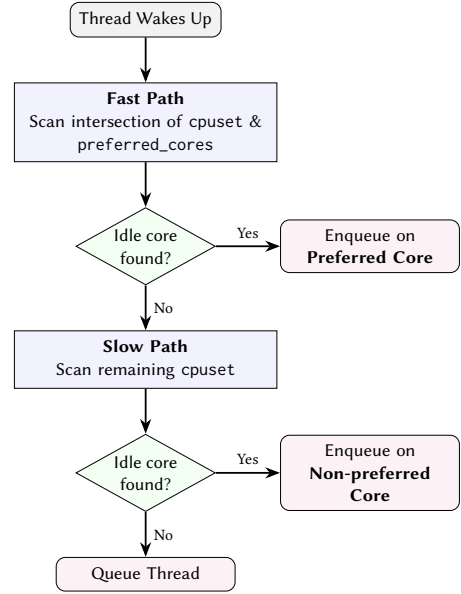
\begin{figure}[htbp]
    \centering
    \resizebox{0.7\columnwidth}{!}{%
    \begin{tikzpicture}[
        node distance=0.7cm and 1cm, %
        font=\footnotesize\sffamily,  %
        startstop/.style={rectangle, rounded corners, align=center, draw=black, fill=gray!10, inner sep=4pt},
        process/.style={rectangle, align=center, text width=3.8cm, draw=black, fill=blue!5, inner sep=4pt},
        decision/.style={diamond, aspect=1.8, align=center, draw=black, fill=green!5, inner sep=1pt},
        action/.style={rectangle, rounded corners, align=center, text width=2.5cm, draw=black, fill=purple!5, inner sep=4pt},
        arrow/.style={thick, ->, >=Stealth}
    ]

    \node[startstop] (wakeup) {Thread Wakes Up};
    
    \node[process, below=of wakeup] (fastpath) {\textbf{Fast Path}\\Scan intersection of \texttt{cpuset} \&\\\texttt{preferred\_cores}};
    
    \node[decision, below=of fastpath] (dec1) {Idle core\\found?};
    
    \node[action, right=of dec1] (enq_pref) {Enqueue on \\ \textbf{Preferred Core}};
    
    \node[process, below=of dec1] (slowpath) {\textbf{Slow Path}\\Scan remaining \texttt{cpuset}};
    
    \node[decision, below=of slowpath] (dec2) {Idle core\\found?};
    
    \node[action, right=of dec2] (enq_std) {Enqueue on \\ \textbf{Non-preferred  Core}};
    
    \node[action, below=of dec2] (queue) {Queue Thread \\};

    \draw[arrow] (wakeup) -- (fastpath);
    \draw[arrow] (fastpath) -- (dec1);
    
    \draw[arrow] (dec1) -- node[above, font=\scriptsize] {Yes} (enq_pref);
    \draw[arrow] (dec1) -- node[right, font=\scriptsize] {No} (slowpath);
    
    \draw[arrow] (slowpath) -- (dec2);
    
    \draw[arrow] (dec2) -- node[above, font=\scriptsize] {Yes} (enq_std);
    \draw[arrow] (dec2) -- node[right, font=\scriptsize] {No} (queue);

    \end{tikzpicture}%
    } %
    \vspace{-0.2cm} %
    \caption{Core-Aware Scheduling (CAS) wakeup decision tree. CAS first scans a container's preferred cores for an idle CPU and falls back to the broader runnable cpuset only when no preferred core is available, preserving work-conservation while biasing placement toward locality.}
    \label{fig:cas_decision_tree}
\end{figure}

\subsection{Demand-Based Dynamic Region Sizing}
The efficacy of soft affinity depends on the sizing and assignment of the Preferred Cores masks. While modern cluster agents,such as those used by Borg, utilize complex machine-learning models to inform machine-level overcommitment capabilities \cite{bashir2021take}, we find that such models produce forecasts at time horizons too long for the sizing of Preferred Cores masks. The peak CPU utilization of individual containers over a 24-hour horizon tends to be near-or-equal to their requested limits, resulting in the same pigeonhole problem described in Section \ref{sec:motivation:overcommit}.

We observed that while the aggregate \textit{requested} CPU limits on Google machines routinely exceed physical capacity, the aggregate \textit{actual} CPU utilization consistently remains well below the hardware's physical capacity. Therefore, we reasoned that we could reasonably size soft affinity regions based on the recent high-percentile CPU utilization of containers.

Specifically, the cluster agent samples a container's CPU utilization every second. Let $u_i$ represent the CPU utilization measured during the 1-second interval $i$. For a trailing 5-minute window ending at time $t$ (consisting of 300 discrete 1-second measurements), we formally define $Demand(t)$ as the $p$-th percentile of this set of observations:
\begin{equation*}
    Demand(t) = \text{Percentile}_{p} \Big( \{ u_{t-\tau} \mid \tau \in [1, 300] \} \Big)
\end{equation*}

By utilizing a high percentile (e.g., $p=99$), this scheme allows us to ensure the soft affinity regions are sized to comfortably match or exceed the actual CPU demand for the vast majority of time, without encountering the pigeonhole problem. We integrated this predictor into our userspace cluster agent daemon, Borglet \cite{verma2015borg}.

\begin{figure}[htbp]
    \centering
    \includegraphics[width=\columnwidth]{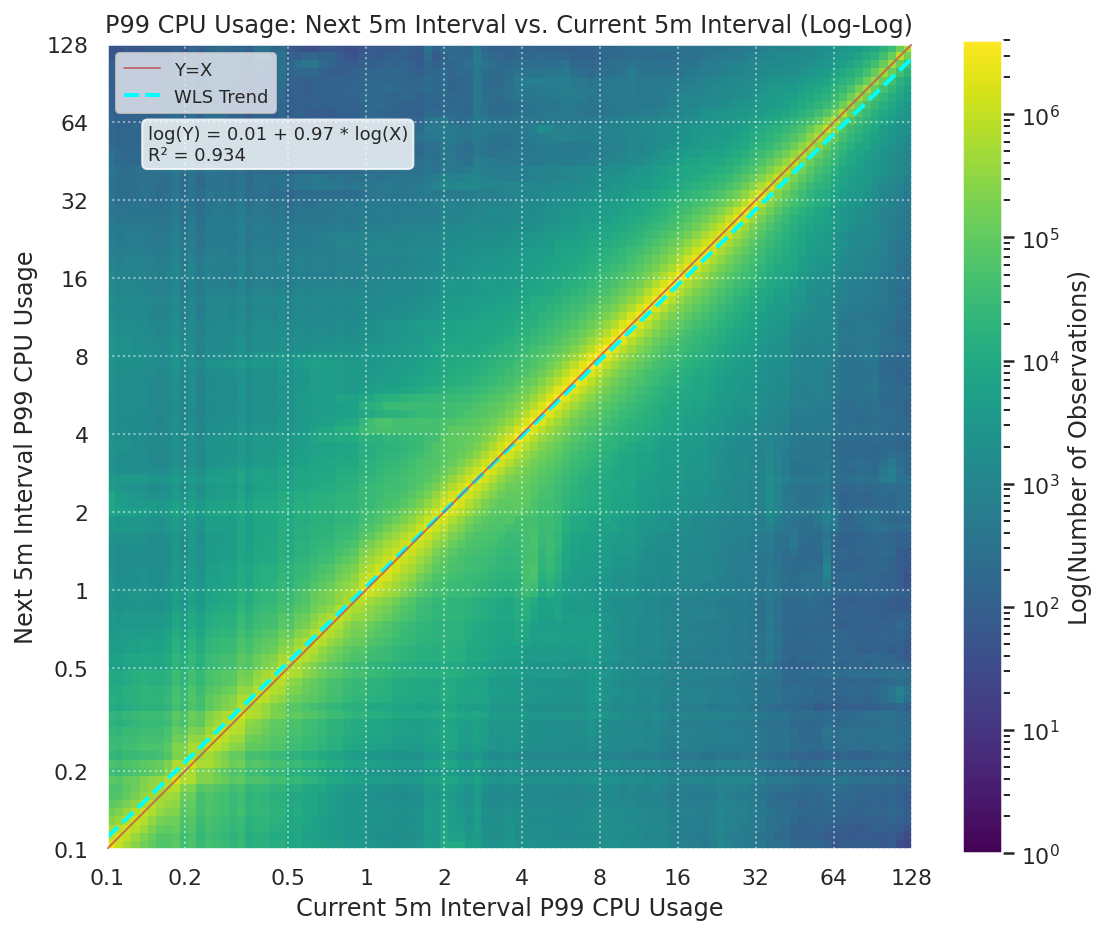}
    \caption{Density map of p99 workload CPU utilization across consecutive 5-minute intervals. The concentration near the $Y=X$ line indicates that recent p99 utilization is strongly predictive of near-term demand ($R^2$ of 0.934).}
    \label{fig:p99_gcu_heatmap}
\end{figure}

Figure \ref{fig:p99_gcu_heatmap} shows the correlation between the trailing 5-minute p99 CPU usage and the subsequent 5-minute p99 CPU usage of containers, indicating that our method projects near-term demand with high accuracy.

\subsubsection{Peak Parallelism Handling}
As seen in Figure \ref{fig:parallelism_heatmap}, datacenter workloads are typically bursty, with peak parallelism that often exceeds their requested limits and average CPU usage. For that reason, while the scheme described above would ensure soft affinity regions are sized to accommodate \textit{average} CPU usage, it would fail to handle microsecond-scale bursts of CPU usage. 

To minimize thread spillover beyond the soft affinity region when applications' parallelism exceeds average usage, we enlarge the assigned soft affinity regions to better load-balance usage across the remainder of the processor. Doing so adds a buffer for applications to spill threads, instead of immediately spilling to shared CPU regions. This scheme is feasible since the sum of \textit{actual} usage rarely exceeds physical capacities; in most of the fleet, we typically see machine utilization below 50\%.  The specific algorithms are discussed in the following section.

\subsection{Topology-Aware Core Allocation}
Using the per-container demand predictions, the Borglet core allocator assigns containers disjoint Preferred Cores while respecting the processor's physical topology. The allocation granularity varies by the dominant source of interference on platform-specific architectures.

\subsubsection{Chiplet-Based / Split-LLC Systems}
We initially designed Algorithm \ref{alg:soft_chiplets} of Affinity Tailor for processor architectures that feature numerous core complexes (chiplets) with their own individual LLC. Data crossing chiplet boundaries must travel through shared memory interconnects, resulting in higher access latencies. The chiplet boundaries impose a steep performance cliff, making it suited as a natural partitioning mechanism to which we aligned our algorithm's allocation units.

Algorithm \ref{alg:soft_chiplets} first schedules workloads with CPU demand under 1 chiplet in ascending order of demand, as single-chiplet assignment obviates the need for inter-chiplet assignment. It then proceeds with scheduling remaining workloads in descending order of demand to minimize fragmentation of large containers.

The algorithm performs an ascending combinatorial search to find the absolute minimum number of chiplets required to satisfy each container's demand. When multiple valid chiplet sets of the same size exist, the algorithm breaks ties by selecting the set with the largest residual capacity, effectively inflating the Preferred Cores regions to provide the most additional capacity for bursting. We find that the combinatorial search is viable, since the number of chiplets per processor remains small, ranging only up to 16 chiplets. A combinatorial search is used due to additional core allocation policy constraints not discussed in this paper.

\subsubsection{Monolithic-LLC Systems}
We later developed Algorithm \ref{alg:simplified_affinity} of Affinity Tailor for processor architectures that feature a monolithic LLC. In this architecture, the performance cliff imposed by LLC boundaries does not exist, since the LLC is shared equally among all cores. We reasoned that the dominant source of interference on these systems are core-local structures such as the L1/L2 caches, branch predictors, and prefetchers. Therefore, we designed the algorithm to have core-granularity allocation units. The design of this algorithm was also driven by the significant core-level effects we observed during the evaluation of the chiplet-based Algorithm \ref{alg:soft_chiplets}. 

Algorithm \ref{alg:simplified_affinity} calculates a per-socket scaling factor, equal to the ratio of available physical cores to the aggregate predicted demand of containers assigned to the socket. This scaling factor is used to enlarge the Preferred Cores regions, providing for a buffer region to absorb bursts of CPU utilization without spilling.

We plan to unify these two algorithms in future work, which will be discussed in Section \ref{sec:discussion}.

\begin{algorithm}
\caption{Affinity Tailor for Split-LLC Systems}
\label{alg:soft_chiplets}
\begin{algorithmic}[1]
\State \textbf{Input:} Set of containers $T$, Set of chiplets $C$
\State \textbf{Output:} Chiplet soft affinity masks for containers

\State $C_{cap}[c] \gets \text{Initial CPU capacity for each chiplet } c \in C$
\State $Cap_{chiplet} \gets \text{CPU capacity of a single chiplet}$

\State $T_{single} \gets \{ t \in T \mid \text{Demand}(t) \le Cap_{chiplet} \}$
\State $T_{multi} \gets T \setminus T_{single}$
\State $\text{SortedT} \gets \textsc{SortAscending}(T_{single})$
\State $\text{SortedT} \gets \text{SortedT} \cup \textsc{SortDescending}(T_{multi})$

\For{\textbf{each} container $t \in \text{SortedT}$}
    \State $d_t \gets \text{Demand}(t)$
    \State $\text{BestSet} \gets \emptyset$
    
    \State // Search for smallest chiplet combos fitting demand
    \For{$i = 1$ \textbf{to} $|C|$}
        \State $\text{Sets}_i \gets$ All combinations of $C$ of size $i$
        \State $\text{ValidSets} \gets \{ s \in \text{Sets}_i \mid \sum_{x \in s} C_{cap}[x] \ge d_t \}$
        
        \If{$\text{ValidSets} \neq \emptyset$}
            \State // Tie-break: Maximize capacity for bursts
            \State $\text{BestSet} \gets \arg\max_{s \in \text{ValidSets}} (\sum_{x \in s} C_{cap}[x])$
            \State \textbf{break}
        \EndIf
    \EndFor
    
    \State \If{$\text{BestSet} \neq \emptyset$}
        \State $\textsc{ApplySoftAffinity}(t, \text{BestSet})$
        \State $\textsc{UpdateCapacities}(C_{cap}, \text{BestSet}, d_t)$
    \EndIf
\EndFor
\end{algorithmic}
\end{algorithm}

\begin{algorithm}
\caption{Affinity Tailor for Monolithic-LLC Systems}
\label{alg:simplified_affinity}
\begin{algorithmic}[1]
\State \textbf{Input:} Set of containers $T$
\State \textbf{Output:} CPU affinity masks for eligible containers

\State $C_{public} \gets$ Identify all non-reserved CPU cores
\State $A_{cores} \gets$ Count of non-reserved cores

\State // Step 1: Compute scaling factor
\State $D_{total} \gets \sum_{t \in T} \text{Demand}(t)$
\State $F_{scale} \gets A_{cores} / D_{total}$

\State // Step 2: Compute and assign cores to each container
\For{each container $t \in T$}
    \State $D_{scaled} \gets \lceil \text{Demand}(t) \times F_{scale} \rceil$
    \State $C_{alloc} \gets$ Pick $D_{scaled}$ cores from $C_{public}$
    \State $C_{public} \gets C_{public} \setminus C_{alloc}$
    \State $\textsc{ApplySoftAffinity}(t, C_{alloc})$
\EndFor
\end{algorithmic}
\end{algorithm}

\section{Evaluation Methodology}\label{sec:methodology}

We deployed Affinity Tailor across Google’s production fleet. Our evaluation baseline is a heavily modified, latency-optimized version of the Linux Completely Fair Scheduler (CFS). While recent work on scheduling architectures frequently leverages userspace dataplanes or kernel-bypass frameworks \cite{kaffes2019shinjuku, ousterhout2019shenango, fried2020caladan, iyer2023concord, jia2024skyloft, lin2024fast, li2024libpreemptible}, deploying such systems in a warehouse-scale setting involves significant friction. Specifically, they require the use of custom runtimes \cite{ousterhout2019shenango,kaffes2019shinjuku,fried2020caladan,fried2024junction}, compiler-level instrumentation \cite{iyer2023concord}, or new hardware features \cite{jia2024skyloft, lin2024fast, li2024libpreemptible}. In contrast, Affinity Tailor is explicitly designed to retain broad compatibility across processor generations and applications. Evaluation against CFS provides the most accurate and realistic comparison with the deployable state-of-the-art in Google's production environment.

Each evaluated machine runs hundreds of live production services, ranging from highly \textit{latency-sensitive} user-facing production applications (e.g., Search, Spanner) to throughput-oriented background workloads. As Affinity Tailor is specifically targeted at latency-sensitive applications, we report results exclusively for this class of workloads.

Our evaluation spans four distinct hardware architectures:
\begin{itemize}
    \item \textbf{Platforms 1, 2, 3} are successive generations of out-of-order multicores featuring split-LLC chiplet topologies.
    \item \textbf{Platform 4} is a recent out-of-order multicore featuring a monolithic LLC.
\end{itemize}

To quantify fleet-wide efficiency, we measure application throughput, which represents the number of requests served by the application per unit of time. We utilize hardware performance counters to capture detailed microarchitectural metrics, specifically LLC references per kilo-instruction (RPKI), LLC misses per kilo-instruction (MPKI), and branch predictor MPKI, in addition to overall system memory bandwidth utilization. We use LLC RPKI as a proxy metric for the combined efficacy of the L1 and L2 data caches.

\section{Evaluation}\label{sec:evaluation}

\subsection{Preferred Core Residency}
\begin{figure}[htbp] 
    \centering
    \definecolor{metric1}{HTML}{0072B2} %
    \definecolor{metric2}{HTML}{D55E00} %
    \definecolor{metric3}{HTML}{56B4E9} %
    \definecolor{metric4}{HTML}{F0E442} %

    \includegraphics{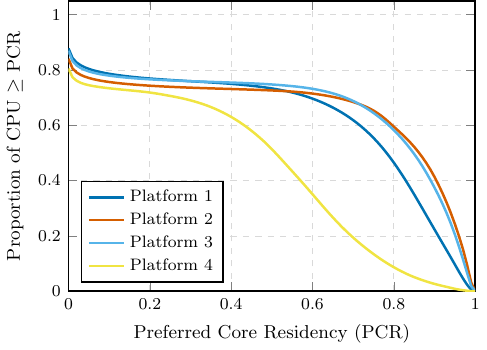}
    \caption{The proportion of total CPU cycles that achieve at least a given Preferred Core Residency (PCR) threshold. Platform 4, which uses Algorithm \ref{alg:simplified_affinity} has lower PCR, indicating that Algorithm \ref{alg:soft_chiplets} is more effective at accommodating bursts.}
    \label{fig:residency_cdf}
\end{figure}

We first validate that Affinity Tailor successfully confines threads to their Preferred Cores. We define \textbf{Preferred Core Residency (PCR)} as the fraction of an application's total CPU time spent executing within its assigned Preferred Cores. Figure \ref{fig:residency_cdf} plots the observed PCR across the evaluated platforms.
$$PCR = \frac{T_{preferred}}{T_{total}}$$
Across the evaluated chiplet-based platforms, we see that 72\% of total CPU time was executed with a PCR exceeding 60\%, while approximately 60\% of CPU time achieved a PCR greater than 80\%. In contrast, the monolithic-LLC Platform 4 exhibited a comparatively lower aggregate PCR. These findings suggest that Algorithm \ref{alg:soft_chiplets} provisions a larger Preferred Cores region for individual applications, thereby better accommodating thread spilling during microsecond-scale bursts than the proportional scaling approach of Algorithm \ref{alg:simplified_affinity}.

\subsection{Application Throughput Impact}\label{sec:eval:fleet}
\begin{figure*}[t]
    \centering

    \definecolor{metric1}{HTML}{1B0C42} %
    \definecolor{metric2}{HTML}{781C6D} %
    \definecolor{metric3}{HTML}{CF4446} %
    \definecolor{metric4}{HTML}{F9A242} %

    \definecolor{tput}{HTML}{332288}    %
    \definecolor{tputGB}{HTML}{88CCEE}  %
    
    \begin{minipage}{0.65\textwidth}
        \centering
        \begin{tikzpicture}
            \begin{axis}[
                width=\textwidth,
                height=5.5cm,
                ybar=2pt, %
                bar width=10pt,
                enlarge x limits=0.2,
                ylabel={\% Change vs. Baseline},
                symbolic x coords={Platform 1, Platform 2, Platform 3, Platform 4},
                xtick=data,
                ymin=-30, ymax=5,
                ymajorgrids=true,
                grid style=dashed,
                area legend,
                legend style={
                    at={(0.5, -0.25)}, %
                    anchor=north,
                    legend columns=-1,
                    draw=none, %
                    fill=none
                },
                nodes near coords,
                visualization depends on={rawy \as \rawy},
                every node near coord/.append style={
                    font=\tiny, 
                    rotate=0,
                    anchor={\ifdim\rawy pt<0pt north\else south\fi}
                },
                nodes near coords align={vertical},
            ]
            
            \addplot[fill=metric1, draw=black!70, point meta=explicit symbolic] coordinates {
                (Platform 1, 0) [N/A*] (Platform 2, 0.1) [0.1] (Platform 3, 0.8) [0.8] (Platform 4, -0.9) [-0.9]
            };
            
            \addplot[fill=metric2, draw=black!70, point meta=explicit symbolic] coordinates {
                (Platform 1, 0) [N/A*] (Platform 2, -7.3) [-7.3] (Platform 3, -24.4) [-24.4] (Platform 4, -1.1) [-1.1]
            };
            
            \addplot[fill=metric3, draw=black!70, point meta=explicit symbolic] coordinates {
                (Platform 1, 0) [N/A*] (Platform 2, -7.3) [-7.3] (Platform 3, -25.6) [-25.6] (Platform 4, -0.3) [-0.3]
            };
            
            \addplot[fill=metric4, draw=black!70] coordinates {
                (Platform 1, -2.2) (Platform 2, -0.5) (Platform 3, -1.8) (Platform 4, -0.6)
            };
            
            \legend{LLC RPKI, LLC MPKI, LLC Miss Rate, Branch MPKI}
            \end{axis}
        \end{tikzpicture}
        \vspace{2pt}
        \raggedright \textit{\scriptsize *We were unable to collect these metrics on Platform 1.}
    \end{minipage}
    \hfill
    \begin{minipage}{0.32\textwidth}
        \centering
        \begin{tikzpicture}
            \begin{axis}[
                width=\textwidth,
                height=5.5cm,
                ybar=2pt, %
                bar width=10pt, %
                enlarge x limits=0.25,
                ylabel={Throughput Uplift (\%)},
                symbolic x coords={Platform 1, Platform 2, Platform 3, Platform 4},
                xtick=data,
                x tick label style={rotate=30, anchor=east, align=right}, %
                ymin=0, ymax=15,
                ymajorgrids=true,
                grid style=dashed,
                area legend, %
                legend style={
                    at={(0.5, -0.35)}, %
                    anchor=north,
                    legend columns=2,
                    draw=none, 
                    fill=none
                },
                nodes near coords,
                every node near coord/.append style={font=\tiny, /pgf/number format/fixed, /pgf/number format/precision=2},
            ]
            
            \addplot[fill=tput, draw=black!70] coordinates {
                (Platform 1, 7.6) (Platform 2, 7.3) (Platform 3, 11.7) (Platform 4, 3.6)
            };
            
            \addplot[fill=tputGB, draw=black!70] coordinates {
                (Platform 1, 4.6) (Platform 2, 4.1) (Platform 3, 7.5) (Platform 4, 2.9)
            };
            
            \legend{per-CPU, per-GB}
            \end{axis}
        \end{tikzpicture}
    \end{minipage}
    
    \caption{Performance impact of Affinity Tailor. The left plot details the effect on cache and branch predictor misses, while the right plot highlights the aggregate application throughput uplift (per-CPU and per-GB). We see significant throughput gains across all evaluated platforms, correlated with improved microarchitectural efficacy.}
    \label{fig:platform_evaluation}
\end{figure*}
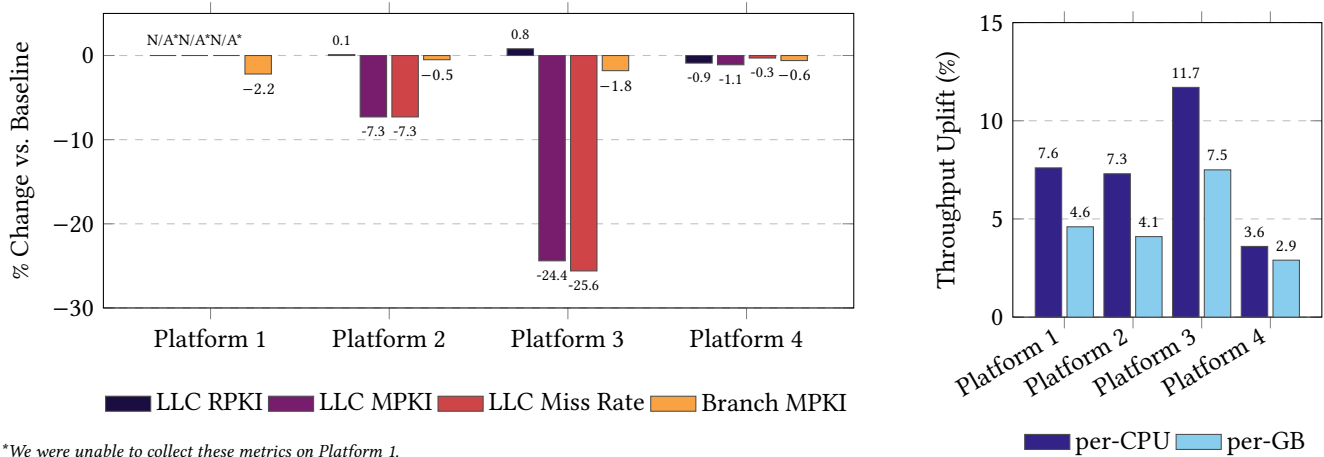
Figure \ref{fig:platform_evaluation} shows that the deployment of Affinity Tailor yields substantial application throughput improvements across all evaluated platforms. We observed aggregate per-CPU throughput improvements ranging from 3.6\% on Platform 4 up to 11.7\% on Platform 3. When evaluating the aggregate throughput impact per-GB of memory, we see parallel gains ranging from 2.9\% to 7.5\%. Faster execution shortens memory residency and reduces memory footprints. Our telemetry reveals that these throughput improvements are correlated with improved microarchitectural efficiency, discussed in the following subsection.

\subsection{Microarchitectural Improvements}
\subsubsection{Core-Level Cache Efficacy}
We evaluate Last-Level Cache References Per Kilo-Instruction (LLC RPKI) as an inverse proxy for L1 and L2 cache hit rates; because a reference is only issued to the LLC when a memory access misses in both core-private caches, a lower LLC RPKI indicates improved L1/L2 efficacy. Across the evaluated platforms, LLC RPKI exhibited varying behavior dictated by the specific Affinity Tailor algorithm in use.

On Platform 4, which utilizes the fine-grained Algorithm \ref{alg:simplified_affinity}, we observe a 0.9\% decrease in LLC RPKI. Algorithm \ref{alg:simplified_affinity} generally provisions smaller, strictly disjoint Preferred Core regions, ensuring that distinct applications rarely interleave on the same physical cores. We reason that the per-application working sets remains tightly bound to disjoint sets of core-private caches, allowing better retention of relevant cache lines without requiring trips to the LLC. 

Conversely, on Platforms 2 and 3—which utilize the chiplet-granularity Algorithm \ref{alg:soft_chiplets}—LLC RPKI stayed flat or increased by up to 0.8\%. Because Algorithm \ref{alg:soft_chiplets} allocates in multiples of whole chiplets, the Preferred Cores regions are larger and shared between more applications. Consequently, it is more likely for threads of distinct applications to be interleaved on the same core. Furthermore, CAS lacks some of the CPU-granularity wakeup affinity mechanisms that are present in CFS.

\subsubsection{Branch Prediction Accuracy}
Across all evaluated platforms, we observed a reduction in Branch Predictor Misses Per Kilo-Instruction (Branch MPKI), ranging from a 0.5\% reduction on Platforms 2, up to a 2.2\% reduction on Platform 1. We believe these reductions arise from improved retention of execution history by the branch predictors, since Affinity Tailor reduces context switching of disparate workloads onto the same physical core. Thus, Affinity Tailor allows execution units to maintain deep, highly accurate and relevant historical state.

\subsubsection{Last-Level Cache Effects}
The impact of Affinity Tailor is most pronounced at the LLC level, particularly on split-LLC systems where chiplet boundaries act as severe performance cliffs. This is evidenced by steep reductions in both LLC MPKI and overall LLC Miss Rates. On Platform 2, LLC miss rates dropped by 7\%, and on Platform 3, they dropped by 26\%. These results are expected, since Algorithm \ref{alg:soft_chiplets} was designed to minimize cross-chiplet sharing of cache lines.

On Platform 4, which has a monolithic LLC, we reason that the 1.1\% and 0.3\% improvement in LLC MPKI and LLC miss rates, respectively, are downstream effects of improved L1/L2 cache efficacy and improved branch predictor accuracy.

\subsubsection{Memory Bandwidth and Prefetcher Activation}
\begin{figure}[htbp]
    \centering
    
    \definecolor{metric1}{HTML}{0072B2} %
    \definecolor{metric2}{HTML}{D55E00} %
    \definecolor{metric3}{HTML}{56B4E9} %
    \definecolor{metric4}{HTML}{F0E442} %

    \begin{tikzpicture}
        \begin{axis}[
            ybar=2pt, %
            width=\columnwidth,
            height=6.5cm,
            bar width=9pt,
            enlarge x limits=0.25, %
            title style={font=\bfseries},
            ylabel={Change in MemBW Util (\%)},
            symbolic x coords={Avg, P90, P99},
            xtick=data,
            label style={font=\small}, 
            tick label style={font=\footnotesize},
            ymin=-5, ymax=1.5,
            ytick={-5, -4, -3, -2, -1, 0, 1},
            ymajorgrids=true,
            grid style={dashed, gray!30},
            extra y ticks={0},
            extra y tick style={grid=major, grid style={solid, thick, black!60}},
            area legend,
            legend style={
                at={(0.5,-0.2)}, 
                anchor=north, 
                legend columns=-1, 
                font=\footnotesize,
                draw=black!20
            },
            thick
        ]
        
        \addplot[fill=metric1, draw=metric1!80!black] coordinates {
            (Avg, -2.48) 
            (P90, -2.94) 
            (P99, -2.00)
        };
        
        \addplot[fill=metric2, draw=metric2!80!black] coordinates {
            (Avg,  0.64) 
            (P90, -0.17) 
            (P99, -0.95)
        };
        
        \addplot[fill=metric3, draw=metric3!80!black] coordinates {
            (Avg, -2.76) 
            (P90, -4.09) 
            (P99, -4.26)
        };
        
        \addplot[fill=metric4, draw=metric4!80!black] coordinates {
            (Avg, -1.40) 
            (P90, -0.87) 
            (P99, -0.92)
        };
        
        \legend{Platform 1, Platform 2, Platform 3, Platform 4}
        
        \end{axis}
    \end{tikzpicture}
    \caption{Change in average, P90, and P99 memory bandwidth utilization percentage under Affinity Tailor. Tail memory bandwidth utilization declines across all evaluated platforms.}
    \label{fig:membw_util_change}
\end{figure}

\begin{figure}[htbp]
    \centering
    
    \definecolor{experimentColor}{HTML}{117733}    %
    \definecolor{controlColor}{HTML}{DDCC77} %
    
    \begin{tikzpicture}
        \begin{axis}[
            ybar=2pt, %
            width=\columnwidth,
            height=6.5cm,
            bar width=14pt,
            enlarge x limits=0.25, %
            title style={font=\bfseries},
            ylabel={Prefetcher Enabled \%},
            symbolic x coords={Platform 1, Platform 2, Platform 3},
            xtick=data,
            label style={font=\small}, 
            tick label style={font=\footnotesize},
            ymin=0, ymax=70, %
            ytick={0, 10, 20, 30, 40, 50, 60, 70},
            ymajorgrids=true,
            grid style={dashed, gray!30},
            area legend,
            legend style={
                at={(0.5,-0.2)}, 
                anchor=north, 
                legend columns=-1, %
                font=\footnotesize,
                draw=black!20
            },
            thick
        ]
        
        \addplot[fill=controlColor, draw=controlColor!70!black] coordinates {
            (Platform 1, 31.4) %
            (Platform 2, 63.4) %
            (Platform 3, 46.5) %
        };
        
        \addplot[fill=experimentColor, draw=experimentColor!70!black] coordinates {
            (Platform 1, 38.2) %
            (Platform 2, 59.7) %
            (Platform 3, 53.4) %
        };
        
        \legend{Affinity Tailor Disabled, Affinity Tailor Enabled}
        
        \end{axis}
    \end{tikzpicture}
    \caption{Impact of Affinity Tailor on hardware prefetcher enablement with Limoncello in-use. Prefetchers remain enabled for a larger fraction of time when Affinity Tailor reduces memory bandwidth utilization.}
    \label{fig:prefetcher_ratio}
\end{figure}

The reductions in LLC misses naturally lead to significantly less main memory access, thereby reducing memory bandwidth utilization. As shown in Figure \ref{fig:membw_util_change}, Affinity Tailor reduces high-percentile (P90 and P99) memory bandwidth utilization across all evaluated platforms. We note that Platform 2 shows an \textit{increase} in average memory bandwidth utilization, and observe that this is an experimental artifact by CPU-load-based dynamic load balancers, which direct additional work to the experiment group due to its improved efficiency.

At Google, we employ Limoncello \cite{jain2024limoncello} to dynamically disable hardware prefetchers when memory bandwidth approaches the saturation threshold, to avoid severe memory access latency increases. Therefore, the aforementioned reduction in memory bandwidth saturation triggers a compounding microarchitectural benefit: sustained hardware prefetcher activation. 

Because Affinity Tailor lowers main memory-bound traffic, it keeps the system below the hardware prefetcher disablement threshold longer, allowing the prefetchers to remain active longer. As illustrated in Figure \ref{fig:prefetcher_ratio}, the percentage of time hardware prefetchers remained enabled increased when memory bandwidth utilization on the platform decreased. Ultimately, this allows the system to reclaim the latency-hiding benefits of hardware prefetchers that are typically lost under high memory bandwidth utilization, thereby improving application performance.

\subsection{Thread Scheduling Latency}
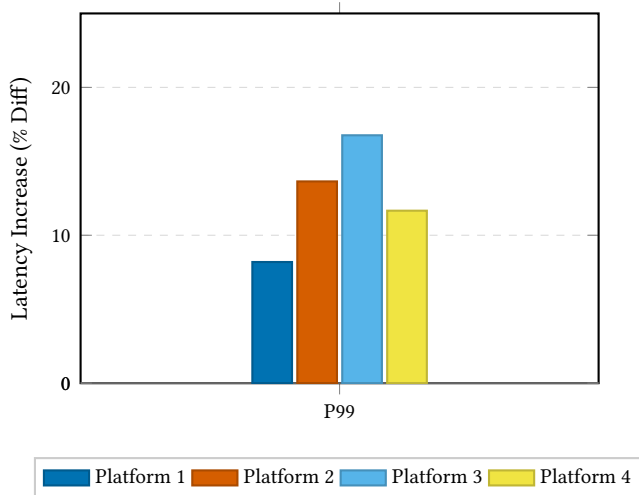
\begin{figure}[htbp]
    \centering

    \definecolor{metric1}{HTML}{0072B2} %
    \definecolor{metric2}{HTML}{D55E00} %
    \definecolor{metric3}{HTML}{56B4E9} %
    \definecolor{metric4}{HTML}{F0E442} %

    \begin{tikzpicture}
        \begin{axis}[
            ybar=2pt, 
            width=\columnwidth,
            height=6.5cm,
            bar width=15pt,          %
            enlarge x limits=0.5,    %
            title style={font=\bfseries},
            ylabel={Latency Increase (\% Diff)},
            symbolic x coords={P99},
            xtick=data,
            label style={font=\small}, 
            tick label style={font=\footnotesize},
            ymin=0, ymax=25,         %
            ymajorgrids=true,
            grid style={dashed, gray!30},
            extra y ticks={0},
            extra y tick style={grid=major, grid style={solid, thick, black!60}},
            area legend,
            legend style={
                at={(0.5,-0.2)}, 
                anchor=north, 
                legend columns=-1, 
                font=\footnotesize,
                draw=black!20
            },
            thick
        ]
        
        \addplot[fill=metric1, draw=metric1!80!black] coordinates {
            (P99, 8.19)
        };
        
        \addplot[fill=metric2, draw=metric2!80!black] coordinates {
            (P99, 13.64)
        };
        
        \addplot[fill=metric3, draw=metric3!80!black] coordinates {
            (P99, 16.76)
        };
        
        \addplot[fill=metric4, draw=metric4!80!black] coordinates {
            (P99, 11.66)
        };
        
        \legend{Platform 1, Platform 2, Platform 3, Platform 4}
        
        \end{axis}
    \end{tikzpicture}
    \caption{A comparison of the percentage increase in thread scheduling latency at the 99th percentile when Affinity Tailor is enabled. We observed significantly increased tail scheduling latencies across all evaluated platforms.}
    \label{fig:latency_percentiles}
\end{figure}

A consequence of Affinity Tailor's current implementation is an increase in thread scheduling latency. By prioritizing the CAS fast-path---which restricts initial thread placement to a container's dynamically sized Preferred Cores---the scheduler bypasses the aggressive load-balancing heuristics present in the baseline CFS that immediately seek out idle cores across the broader socket. Because CAS currently lacks these mature latency optimizations, threads frequently experience localized queueing while waiting for a preferred core to become available, rather than being instantly migrated to a remote idle core. Threads that are scheduled outside their Preferred Cores are also aggressively migrated back to their Preferred Cores.

Figure \ref{fig:latency_percentiles} illustrates the percentage increase in P99 scheduling latency (the time a thread spends in the runnable state waiting for CPU dispatch). As expected from the absence of aggressive idle-search optimizations, the deployment of Affinity Tailor causes substantial latency regressions in the tail. Across all evaluated platforms, we observe 99th percentile scheduling latencies increasing by as much as 17\%.

In traditional operating-system design, as exemplified by CFS and EEVDF, increases in thread-queueing delay are typically avoided, as it is assumed to directly degrade application performance. However, our fleet-wide evaluation reveals that despite an "undesirable" increase in scheduling latency, overall application throughput significantly improves (as detailed in Section \ref{sec:eval:fleet}).

\section{Related Work}\label{sec:related}
\textbf{Locality-Aware System Design:}
Affinity Tailor builds upon a history of chiplet-aware design practices. The Nest scheduler \cite{lawall2022scheduling} concentrates threads onto a dense set of warm cores to exploit higher turbo frequencies, thus improving performance and energy efficiency in lightly-loaded environments. It does not incorporate per-application demand forecasting or target overcommitted multi-tenant environments. Recent proposals like Cache Aware Scheduling \cite{chen2024cas} seek to implement LLC affinity directly within the mainline Linux kernel's load balancer, using purely kernel load metrics without additional context from userspace. Zhou et al. \cite{zhou2024characterizing} showed that sharding TCMalloc's transfer caches into chiplet-local structures improved application throughput.

\textbf{Extensible Scheduling Frameworks:}
There have been many recent efforts to modularize the Linux kernel's scheduling policies, which would allow for faster implementation of Affinity Tailor's Preferred Cores policy. The \texttt{sched\_ext} framework \cite{sched_ext2024} was merged into Linux 6.12, enabling the use of dynamically-loaded BPF scheduling policies. ghOSt \cite{humphries2021ghost} delegates scheduling to userspace agents with shared-memory queues. Enoki \cite{miller2024enoki} supports rapid kernel scheduler development, and Plugsched \cite{ma2023efficient} decouples the Linux scheduler into a loadable module.

\textbf{Resource Partitioning and Throttling:}
Prior systems have extensively studied shared resource contention through the lens of partitioning and throttling. $CPI^2$ \cite{zhang2013cpi2} utilizes hardware performance counters to continuously monitor, detect, and throttle antagonistic workloads. Heracles \cite{lo2015heracles} employs feedback controllers to safely isolate latency-critical workloads from best-effort background workloads by dynamically managing CPU, memory, and network resources. PARTIES \cite{chen2019parties} introduced QoS-aware partitioning of LLC ways, memory bandwidth, and cores for multiple services. CLITE \cite{patel2020clite} and OLPart \cite{chen2023olpart} improved upon PARTIES with Bayesian optimization and online learning, respectively. These systems govern shared resource allocation (cache ways, memory bandwidth, time-on-core); Affinity Tailor governs thread-to-core placement.

\section{Discussion and Future Directions}\label{sec:discussion}
\subsection{Unifying Affinity Tailor Algorithms}
Our evaluation of Affinity Tailor revealed a notable divergence in microarchitectural behavior between our two Preferred Cores allocation algorithms. Algorithm \ref{alg:soft_chiplets} (chiplet-\hspace{0pt}granularity) successfully mitigated cross-chiplet LLC thrashing on split-LLC architectures (Platforms 2 and 3). However, it yielded minimal improvements in core-private L1/L2 cache hit rates, as evidenced by stagnant to increased LLC RPKI. Conversely, Algorithm \ref{alg:simplified_affinity} (core-granularity) deployed on the monolithic-LLC Platform 4 demonstrated strong L1/L2 cache efficacy improvements by tightly packing threads onto disjoint sets of CPUs.

This dichotomy highlights a clear path for future optimization: unifying the two algorithms into a tiered, hierarchical allocation algorithm. For massively parallel, chiplet-based processors, the cluster agent should first bin applications into the optimal minimum set of chiplets to minimize cross-chiplet and main-memory-bound traffic. Subsequently, \textit{within} those selected chiplets, the agent should apply the fine-grained, demand-scaled core allocation logic to better govern core-level predictive state sharing. By combining these approaches, future systems may simultaneously harvest the LLC improvements of coarse-grained chiplet-based partitioning and the deep, core-level state preservation of fine-grained partitioning.

\subsection{Towards Locality-Aware Scheduling}
Perhaps the most profound architectural insight from the deployment of Affinity Tailor is the observation that aggressively migrating threads to minimize immediate queueing is a flawed optimization target in hyperscale datacenters built on hardware with deeply stateful microarchitectures. Our findings provide compelling evidence that a strict adherence to work-conservation—a foundational principle in modern schedulers like CFS and EEVDF—can be counterproductive on modern hardware and software architectures.

Currently, Affinity Tailor implements improved spatial locality via a cooperative hardware-software architecture: a userspace daemon calculates the boundaries, and the kernel enforces the Preferred Cores soft affinity fast-path through CAS. However, these results suggest that operating system schedulers could benefit from a more fundamental evolution. Future OS schedulers may need to natively treat spatial locality and microarchitectural state ``warmth'' as first-class scheduling parameters. Such schedulers could dynamically weigh the microsecond penalty of localized queueing against the latency costs of remote memory fetches, cold branch predictors, and degraded hardware prefetchers. Shifting the core scheduling paradigm from aggressive load-balancing to a dynamic, locality-aware approach represents a promising frontier in datacenter efficiency.

Conversely, these results suggest a parallel evolution path in hardware design. Because deeply stateful predictors are vulnerable to frequent context switching, future architectures might benefit from explicitly accommodating these OS scheduling realities. Hardware designers could explore mechanisms for rapid state recovery rather than presuming prolonged, consistent execution of applications.

\subsection{Cross-Stack Layering for Scheduling}
While Affinity Tailor demonstrates the value of coordinating userspace demand prediction with kernel-enforced soft affinity, its current implementation applies spatial locality heuristics uniformly across all configured workloads. Operating systems, functioning purely at the hardware abstraction layer, lack semantic understanding of the applications they schedule. Consequently, the kernel alone is ill-equipped to determine which workloads actually benefit from microarchitectural state preservation.

Future scheduling architectures can embrace deeper cross-stack layering, leveraging the rich contextual metadata accessible to userspace cluster agents like Borglet. The agent possesses comprehensive knowledge of a job's planned resource constraints, historical execution phases, and overall workload archetype. These signals can directly inform thread-scheduling decisions.

For instance, a lightweight RPC forwarding service is primarily network I/O bound; tightly packing its threads into a constrained set of Preferred Cores may artificially bottleneck its processing throughput without yielding particularly meaningful cache-hit improvements. The agent could proactively classify workloads by their affinity-sensitivity, supplying these semantic hints to Affinity Tailor. In turn, Affinity Tailor can restrict spatial locality heuristics to state-heavy, compute-bound applications, while allowing stateless or network-bound microservices to aggressively load-balance across the socket. This targeted application of spatial locality would allow overcommitted systems to more effectively apportion scarce resources.

\section{Conclusion}\label{sec:conclusion}
As hyperscalers aggressively overcommit hardware, conventional operating system schedulers inadvertently exacerbate microarchitectural interference by prioritizing work-conservation, constantly dispersing workloads across massive processor topologies and destroying core-local predictive state. Affinity Tailor is a dynamic scheduling architecture utilizing a novel Preferred Cores kernel mechanism to establish permeable, per-application, soft affinity boundaries sized to near-term demand forecasts. Deployed across Google's global production fleet, Affinity Tailor confines execution to warm microarchitectural domains while still permitting bursts onto external cores, improving aggregate application throughput by up to 12\% per-CPU and 7\% per-GB memory. Ultimately, our deployment demonstrates that the microsecond-scale queueing penalty incurred by waiting for a preferred core is eclipsed by the execution speedups gained from inheriting a hot microarchitectural environment, suggesting that future operating system designs should consider prioritizing spatial locality over strict work-conservation.

\begin{acks}
Affinity Tailor was made possible by numerous Google engineering teams across many years. We would like to especially thank Yiyan Lin and Sundar Dev for their early work in productionizing Preferred Cores. We also thank Xiangling Kong, Nan Deng, Zhiyuan Liu, Li Li, Dagang Wei, Trang Tran, Sahil Shekhawat, Shiyu Hu, Peilin Ye, Corentin Pescheloche, Steve Zekany, Colin Rioux, Darryl Gove, Aleksei Shchekotikhin, Nilay Vaish, Patrick Xia, Wanying Lu, Tae Jun Ham, Haiming Liu, and Akanksha Jain for their guidance and support. Furthermore, we are grateful to Sotiris Apostolakis, Tipp Moseley, and Sree Kodakara for their invaluable feedback. Gemini was utilized to generate sections of this Work, including text, figures, and citations.
\end{acks}

\bibliographystyle{ACM-Reference-Format}
\bibliography{references}

\end{document}